\documentclass[12pt]{JHEP3}

\usepackage{amssymb}
\usepackage{graphics}
\usepackage{epsfig}
\usepackage{graphicx}
\usepackage{subfigure}
\usepackage{bm}

\def\beq{\begin{equation}}
\def\eeq{\end{equation}}
\def\baq{\begin{eqnarray}}
\def\eaq{\end{eqnarray}}

\def\k{{\bf k}}
\def\q{{\bf q}}

\def\G{{\rm G}}

\def\fnl{f_{\rm NL}}
\def\nfnl{n_{f_{\rm NL}}}
\def\gnl{g_{\rm NL}}
\def\ngnl{n_{g_{\rm NL}}}
\def\taunl{\tau_{\rm NL}}
\def\ntaunl{n_{\tau_{\rm NL}}}
\def\so{\sigma_{\rm osc}}
\def\hso{\hat{\sigma}_{\rm osc}}

\title{Strongly scale-dependent polyspectra from curvaton self-interactions}

\author{Christian T.~Byrnes$^a$,
Kari Enqvist$^{b}$, Sami Nurmi$^{c}$,
Tomo Takahashi$^{d}$ \\
$^a$ Fakult{\"a}t f{\"u}r Physik, Universit{\"a}t Bielefeld,
Postfach 100131, 33501 Bielefeld, Germany;\\
$^b$Physics Department, University of Helsinki, and Helsinki Institute of Physics,
FIN-00014 University of Helsinki, Finland;\\
$^c$NORDITA, SE-106 91, Stockholm, Sweden;\\
$^d$Department of Physics, Saga University, Saga 840-8502, Japan }

\abstract {We study the scale dependence of the non-linearity parameters
$\fnl$ and $\gnl$ in curvaton models with self-interactions. We show
that the spectral indices $\nfnl={\rm d}\,{\rm ln}|\fnl|/{\rm
d}\,{\rm ln}\,k$ and $\ngnl={\rm d}\,{\rm ln}|\gnl|/{\rm d}\,{\rm
ln}\,k$ can take values much greater than the slow--roll
parameters and the spectral index of the power spectrum. This means
that the scale--dependence of the bi and trispectrum could be easily
observable in this scenario with Planck, which would lead to tight
additional constraints on the model. Inspite of the highly
non-trivial behaviour of $\fnl$ and $\gnl$ in the curvaton models
with self-interactions, we find that the model can be falsified if
$\gnl(k)$ is also observed.}


\preprint{BI-TP 2011/24 \\ HIP-2011-22/TH \\ NORDITA-2011-66}

\begin{document}

\section{Introduction}

Non-Gaussianity of the primordial perturbations can efficiently
discriminate between different models of inflation. It is by now
well known that both the strength and statistical properties of
primordial non-Gaussianities depend crucially on the details of the
inflationary model. While conventional slow roll models of inflation
with canonical dynamics typically predict negligible
non-Gaussianity, non-minimal constructions may generate observable
non-Gaussianity. For a selection of recent reviews see \cite{Liguori:2010hx,Chen:2010xka,Byrnes:2010em,Komatsu:2010hc,Wands:2010af}. Examples of such scenarios include models with
non-canonical kinetic terms or a breakdown of slow roll dynamics and
models where the primordial perturbations are generated at the end
of inflation, like in modulated reheating \cite{Dvali:2003em,Kofman:2003nx}, or after the end of
inflation, like in the curvaton scenario \cite{curvaton,LUW}.

The simplest type of non-Gaussianity is the so called local form.
The local Ansatz for the primordial curvature perturbation reads
\beq
  \label{local}
  \zeta_{\k}=\zeta^{\G}_{\k}+\frac{3}{5}\fnl(\zeta^{\G}\star\zeta^{\G})_{\k}+
  \frac{9}{25}\gnl(\zeta^{\G}\star\zeta^{\G}\star\zeta^{\G})_{\k}
  \ ,
  \eeq
where $\zeta^{\G}$ is a Gaussian field and the non-linearity
parameters $\fnl$ and $\gnl$ are constants. The star denotes a
convolution. In \cite{Byrnes:2009pe,Byrnes:2010ft} it was shown that non-linearities of the
field equations in general give rise to a mild scale dependence of
$\fnl$ and $\gnl$ even in models with canonical slow-roll dynamics
during inflation, which typically have been analyzed using the local
Ansatz. Deviations from the local Ansatz were analyzed previously
from an observational point of view in \cite{LoVerde:2007ri,Sefusatti:2009xu} and more recently
in \cite{Shandera:2010ei,Becker:2010hx}. The scale dependence of $\fnl$ and $\gnl$ can be
described by the quantities
  \beq
  \label{defnl}
  \nfnl=\frac{{\rm d}\,{\rm ln}|\fnl|}{{\rm d}\,{\rm ln}\,k}\ ,\qquad
  \ngnl=\frac{{\rm d}\,{\rm ln}|\gnl|}{{\rm d}\,{\rm ln}\,k}\ ,
  \eeq
and in models with slow roll dynamics during inflation the typical
magnitude of $\nfnl$ and $\ngnl$ is set by slow roll parameters
\cite{Byrnes:2010ft}. Hence the local Ansatz (\ref{local}) gets
replaced by a quasi-local form with $\fnl$ and $\gnl$ being weakly
$k$-dependent functions. Although the scale dependence is typically
weak (for an exception see \cite{Riotto:2010nh}), it could be an
observable effect which makes this topic very interesting. This
topic is also interesting in models of non--local Gaussianity, see
e.g.~\cite{Chen:2005fe,Bartolo:2010im,Noller:2011hd,Burrage:2011hd}.

In this work we compute the scale dependency of the non-linearity
parameters in self-interacting curvaton models. The scale dependence
in the limit of weak interactions, where the self-interaction does not
dominate over the quadratic part in the curvaton potential, was
considered already in \cite{Byrnes:2010xd}. Here we analyze also the
self-interaction dominated regime, which turns out to be
particularly interesting. In this regime the non-linearity
parameters $\fnl$ and $\gnl$ depend sensitively on the curvaton
value at the time of inflation
\cite{Enqvist:2009ww,Kawasaki:2011pd,Enqvist:2005pg,Enqvist:2008gk,Enqvist:2009zf}.
In particular, they oscillate around the naive estimates $|\fnl|\sim
r_{\rm dec}^{-1}$ and $|\gnl|\sim r_{\rm dec}^{-2}$, where $r_{\rm
dec}$ denotes the curvaton contribution to the total energy density
at the decay time. We find that the non-linearity parameters become
strongly scale-dependent in the regions $|\fnl|\ll r_{\rm dec}^{-1},
|\gnl|\ll r_{\rm dec}^{-2}$. For small values of $r_{\rm dec}$ the
non-Gaussian effects in these regions are at the observable level
and the scale dependence will be a detectable feature of this model in
the near future. We also comment on how the results could be
generalized to any models where the primordial perturbation arises
from a component which is subdominant during inflation.

The paper is organized as follows: in Section \ref{sec2} we review
the self-interacting curvaton scenario to the extent needed in our
analysis. In Section \ref{sec3} we derive expressions for the scale
dependent non-Gaussianity in the curvaton scenario and discuss their
generic features. In Section \ref{sec4} we discuss the case of
quartic self--interactions, deriving both analytical and numerical
results. In Section \ref{sec5}, we discuss the scale dependence in
curvaton models with non-renormalizable self-interactions, while in
Section \ref{sec:interactiondominated} we study the regime of very
large self interactions and find an analytical approximation.
Finally, we present our conclusions in Section \ref{sec6}. We use
the units $M_{\rm P}=(8\pi G)^{-1/2} = 1$ throughout the paper.

\section{Self-interacting curvaton model}
\label{sec2}

We consider self-interacting curvaton models where the curvaton
potential is given by
  \beq
  \label{V}
  V=\frac{1}{2}m^2\sigma^2+\lambda\sigma^{n}\ ,
  \eeq
where $n=4,6,8$.

While the curvaton should oscillate before decaying, and the
quadratic part will typically be dominant at this stage due to the
small field value, the self-interacting part may play an important
role at earlier stages, crucially affecting the predictions of the
curvaton scenario
\cite{Enqvist:2009ww,Kawasaki:2011pd,Enqvist:2005pg,Enqvist:2008gk,Enqvist:2009zf,
Choi:2010re,Fonseca:2011iz,Huang:2008zj,Dimopoulos:2003ss,Kawasaki:2008mc,Chingangbam:2009xi}.
We assume the primordial perturbation arises solely from the
fluctuations of the curvaton field $\sigma$ and neglect the inflaton
contribution. Mixed scenario's were considered in \cite{mixed}.
After the end of inflation the inflaton decays into radiation which
dominates the universe. We assume the curvaton decays
instantaneously into radiation at $H_{\rm dec}=\Gamma$, for a
discussion on the accuracy of this approximation see
\cite{Malik:2006pm,Sasaki:2006kq}. If the curvaton is coupled to
other scalars, it may also decay non-perturbatively through a
parametric resonance \cite{Enqvist:2008be,Chambers:2009ki}, we will not
consider this possibility here.

Using the $\delta N$ formalism \cite{starob85,ss1,Sasaki:1998ug,lms,lr}, the curvature
perturbation can be expressed in the form
  \baq
  \nonumber
  \label{zeta_def}
  \zeta_{\k}&=&N'(t_k)\delta\sigma_{\k}(t_k)+\frac{1}{2}N''(t_k)(\delta\sigma\star\delta\sigma)_{\k}(t_k)
  +\frac{1}{6}N'''(t_k)(\delta\sigma\star\delta\sigma\star\delta\sigma)_{\k}(t_k)+\ldots\\\nonumber
  &=&\frac{2\,r_{\rm dec}}{3}\frac{\sigma_{\rm osc}'}{\sigma_{\rm
  osc}}\delta\sigma_{\k}(t_k)+\frac{r_{\rm dec}}{3}\left(\frac{\sigma_{\rm osc}''}{\sigma_{\rm
  osc}}+\left(\frac{\sigma_{\rm osc}'}{\sigma_{\rm
  osc}}\right)^2\right)(\delta\sigma\star\delta\sigma)_{\k}(t_k)\\
  &&+\frac{r_{\rm dec}}{9}\left(\frac{\sigma_{\rm osc}'''}{\sigma_{\rm
  osc}}+3\frac{\sigma_{\rm osc}''\sigma_{\rm osc}'}{\sigma_{\rm
  osc}^2}\right)(\delta\sigma\star\delta\sigma\star\delta\sigma)_{\k}+\cdots\,,
  \eaq
where the convolutions are defined by
$(\delta\sigma\star\delta\sigma)_{\k}(t_k)=(2\pi)^{-3}\int {\rm
d}\q\,\delta\sigma_{\q}(t_k)\delta\sigma_{\k-\q}(t_k)$. $N(t_k)$
denotes the number of e-foldings from an initial spatially flat
hypersurface $t_k$, corresponding to the horizon exit of the mode
$k$, to some final uniform energy density surface after the decay of
the curvaton. The primes denote derivatives with respect to
$\sigma(t_k)$. $\sigma_{\rm osc}$ sets the scale of the curvaton
envelope during the final quadratic oscillations before the decay,
$\bar{\sigma}(t)=\sigma_{\rm osc}/(mt)^{3/4}$, see equation
(\ref{def_so}) below. $r_{\rm dec}$ measures the curvaton
contribution to the total energy density at the time of decay,
  \beq
  r_{\rm dec}=\frac{3}{4}\frac{\rho_{\sigma}}{3H^2}\Big|_{\rm dec}\simeq \frac{1}{2\sqrt{2}}\,\sigma_{\rm
  osc}^2\left(\frac{m}{\Gamma}\right)^{1/2}\ ,
  \eeq
and the results are computed to first order in $r_{\rm dec}$
throughout this work.

The information about curvaton self-interactions in (\ref{zeta_def})
is essentially encoded into the derivatives of $\sigma_{\rm osc}$.
In general, it is not possible to compute $\sigma_{\rm
osc}(\sigma(t_k))$ analytically. However, we can obtain some generic
information by just looking at the evolution equation for the
curvaton field in the radiation dominated epoch,
  \beq
  \ddot{\sigma}+\frac{3}{2t}\dot{\sigma}+m^2\sigma+n\lambda\sigma^{n-1}=0\
  .
  \eeq
Switching to the variable $x=mt$ and writing
$\sigma=\sigma(t_k)\xi(x) x^{-3/4}$, we obtain
  \beq
  \label{xi}
  \frac{{\rm d}^2\xi}{{\rm
  d}x^2}+\xi(1+\frac{3}{16}x^{-2})+\frac{ns}{2}\,\xi^{n-1}x^{-3(n-2)/4}=0\ .
  \eeq
Here we have defined a (relative) self-interaction strength parameter $s$ by
  \beq
  \label{s}
  s=\frac{2\,\lambda\sigma(t_k)^{n-2}}{m^2}\ ,
  \eeq
which is simply the ratio of potential energies stored in the
curvaton self-interactions and the bare mass part in (\ref{V}). Beware that various definitions of $s$ and $r_{\rm dec}$ have been used in the literature.

In the asymptotic limit $x\rightarrow \infty$, corresponding to the
regime of quadratic oscillations, (\ref{xi}) has a solution
$\xi_{\rm as}=\hso(s)\,{\rm sin}(x+\varphi(s))$, and we obtain the
asymptotic result
  \beq
  \label{def_so}
  \sigma_{\rm as}(t)=\sigma(t_k) \hso(s) \frac{{\rm
  sin}(mt+\varphi(s))}{(mt)^{3/4}}\equiv \sigma_{\rm osc}\frac{{\rm
  sin}(mt+\varphi(s))}{(mt)^{3/4}}\ .
  \eeq
From this we learn that $\sigma_{\rm osc}$ can be expressed in the
form
  \beq
  \label{sigma_osc(s)}
  \sigma_{\rm osc}(\sigma(t_k))= \sigma(t_k)\hso(s)\ .
  \eeq

\section{Scale-dependent non-Gaussianity}
\label{sec3}

We analyze the scale-dependence of the non-linearity parameters
using the formalism developed in \cite{Byrnes:2009pe,
Byrnes:2010ft}. The curvaton perturbations at horizon crossing,
$\delta\sigma_{\k}(t_k)$, are assumed to be Gaussian. We study the effect of relaxing this assumption in Appendix \ref{appendix}. We denote the
Gaussian part of the curvature perturbation (\ref{zeta_def}) by
  \beq
  \label{zeta_G}
  \zeta_{\k}^{\G}\equiv N'(t_k)\delta\sigma_{\k}(t_k)\equiv\frac{2\,r_{\rm
  dec}}{3}z(s)\frac{\delta\sigma_{\k}(t_k)}{\sigma(t_k)} \ ,
  \eeq
where the function $z(s)$ is given by
  \beq
  \label{z}
  z(s)=\frac{\sigma(t_k)\so'}{\so}=1+\frac{(n-2)s}{\hso}\frac{\partial\hso}{\partial s_{~~~}}\ .
  \eeq
The expression (\ref{zeta_def}) for the curvature perturbation can
now be written as
  \beq
  \label{zeta_k}
  \zeta_{\k}=\zeta_{\k}^{\G}+\frac{3}{5}\fnl(k)(\zeta^{\G}\star\zeta^{\G})_{\k}+\frac{9}{25}
  \gnl(k)(\zeta^{\G}\star\zeta^{\G}\star\zeta^{\G})_{\k}+\cdots\ ,
  \eeq
where the non-linearity parameters are given by
  \baq
  \label{fexps}
  \fnl(k)&=&
  \frac{5}{6}\frac{N''(t_k)}{N'(t_k)^2}=\frac{5}{4 r_{\rm dec}}\left(1+\frac{\so''\so}{\so'{}^2}\right)
  \equiv\frac{5}{4 r_{\rm dec}}\,f(s)\ ,
  \\
  \label{gexps}
  \gnl(k)&=&\frac{25}{54}\frac{N'''(t_k)}{N'(t_k)^3}=\frac{25}{24 r_{\rm dec}^2}
  \left(\frac{\so'''\so^2}{\so'{}^3}+\frac{3\so''\so}{\so'{}^2}\right)\equiv\frac{25}{24 r_{\rm
  dec}^2}\,g(s)\ ,
  \eaq
to leading order in $r_{\rm dec}$. Since we assume $\zeta$ is
generated by a single field, the third non-linearity parameter
$\taunl$, describing the trispectrum together with $\gnl$, is
uniquely determined by $\fnl$
  \[
  \taunl=\left(\frac{6}{5}\fnl\right)^2\ .
  \]
It hence trivially follows that $n_{\tau_{\rm NL}}=2\nfnl$, as is
always the case for a model where $\zeta$ is generated by a
single-source \cite{Byrnes:2010ft}. For a general discussion of the
relation between the (local) non-linearity parameters see
\cite{Suyama:2010uj}.

Information about the curvaton interactions is encoded into the
functions $f(s)$ and $g(s)$, which depend only on the
self-interaction strength parameter $s$. For a purely quadratic
model, $s=0$, they read $f=1$ and $g=0$ since $\sigma_{\rm
osc}\propto\sigma(t_{k})$. In the presence of self-interactions,
$f(s)$ and $g(s)$ become non-trivial functions oscillating between
positive and negative values \cite{Enqvist:2009ww}. The level of
non-Gaussianity may therefore strongly deviate from the the naive
estimates of $|\fnl|\sim r_{\rm dec}^{-1}$ and $|\gnl|\sim r_{\rm
dec}^{-2}$.

Equations (\ref{fexps}) and (\ref{gexps}) are evaluated at the
horizon crossing time $t_{k}$ of the mode $k$ under consideration, which
in general makes $\fnl$ and $\gnl$ scale-dependent
\cite{Byrnes:2009pe, Byrnes:2010ft}. The scale dependence can be
described by the parameters $\nfnl$ and $\ngnl$, which measure
logarithmic derivatives of the non-linearity parameters
(\ref{defnl}). Applying the results of \cite{Byrnes:2009pe,
Byrnes:2010ft} to the curvaton scenario, we find
  \baq
  \label{nfnl_single}
  \nfnl&=&\frac{1}{2}\,\ntaunl=\frac{N'}{N''}
  \frac{V'''}{3H^2}\\\nonumber
  &=&\frac{\eta_{\sigma}}{f(s)}\left(
  \frac{n(n-1)(n-2)s}{z(s)(2+n(n-1)s)}\right)\ ,\\
  \label{ngnl_single}
  \ngnl&=&3\frac{N''^2}{N'''N'}\,\nfnl+\frac{N'}{N'''}
  \frac{V''''}{3H^2}\\\nonumber
  &=&\frac{\eta_{\sigma}}{g(s)}\left(\frac{n(n-1)(n-2)s}{z(s)(2+n(n-1)s)}\left(3f(s)+\frac{n-3}
  {z(s)}\right)\right)\ .
  \eaq
The results are derived to leading order in slow roll, see
\cite{Byrnes:2009pe, Byrnes:2010ft} for details. The functions
$z,f,g$ are defined by (\ref{z}), (\ref{fexps}) and (\ref{gexps}).
The slow--roll parameter $\eta_{\sigma}$ is defined as usual,
$\eta_{\sigma}={V''}/{(3H^2)}={m^2}(2+n(n-1)s)/({6H^2})$. In the
curvaton scenario, the scale dependence of $\fnl$ and $\gnl$ is
entirely generated by curvaton interactions. For a purely quadratic
model the scale dependence vanishes $\nfnl=\ngnl=0$ (note that also
$r_{\rm dec}^2 \gnl = 0$ in this case, subleading corrections in
$r_{\rm dec}$ lead to $\gnl={\cal O}(\fnl)$) as the equation of
motion for $\sigma$ is fully linear to leading order in $r_{\rm
dec}$ \cite{Byrnes:2009pe}. The first detailed study of the scale
dependence of $g_{\rm NL}$ for isocurvature models was made in
\cite{Huang:2011di}.

The results given here are valid for $|\nfnl|$ and $|\ngnl|$ much less than unity \cite{Byrnes:2009pe, Byrnes:2010ft}. To discuss stronger
scale-dependence, the formalism needs to be modified to account for
the non-Gaussianity of the curvaton perturbations
$\delta\sigma_{\k}(t_k)$. A more detailed discussion on this issue
is presented in Appendix \ref{appendix}.

\subsection{Regimes of enhanced scale-dependence}

The complicated dynamics of the self-interacting curvaton scenario
can lead to a considerable enhancement of the scale-dependence. In
the regions where $|f(s)|\ll 1$ or $|g(s)|\ll 1$, the spectral
indices
  \beq
  \nfnl\propto \frac{\eta_{\sigma}}{f(s)}\ ,\qquad \ngnl\propto \frac{\eta_{\sigma}}{g(s)}
  \eeq
can become much larger than the slow-roll scale $\eta_{\sigma}$. The
amplitudes $\fnl\propto f(s)/r_{\rm dec}$ and $\gnl\propto
g(s)/r_{\rm dec}^2$, on the other hand, depend not only on the
self-interaction strength $s$ but also on $r_{\rm dec}$, measuring
the curvaton energy density at the time of its decay. As $r_{\rm
dec}$ can be varied independently of $s$, the non-linearity
parameters $\fnl$ and $\gnl$ can be large even if $f$ or $g$ are
suppressed.

According to \cite{Sefusatti:2009xu}, Planck should be able to probe
the scale-dependence of $\fnl$ to the precision
  \beq
  \Delta\nfnl \simeq 0.1\frac{50}{\fnl}\frac{1}{\sqrt{f_{\rm sky}}}\
  ,
  \eeq
where $f_{\rm sky}$ stands for the fraction of sky observed and the
result is derived taking $\fnl=50, \nfnl=0$ as fiducial values in
the analysis. For CMBpol, the error is expected to be smaller by a
factor of two. Since the error $\Delta\nfnl$ is inversely
proportional to $\fnl$, it is also interesting to consider the
combination $\fnl\nfnl \propto \eta_{\sigma}/{r_{\rm dec}}$ for
which this dependence drops out. The combination $\fnl\nfnl$ is
parameterically suppressed for $\eta_{\sigma}\ll r_{\rm dec}$ but
can become observable for $\eta_{\sigma}\gtrsim r_{\rm dec}$. This
is in accordance with our finding that the regions $|f(s)|\ll 1$,
with $f(s)$ defined in equation (\ref{fexps}), are characterized by
an enhanced scale-dependence. Indeed, the lower bound on $r_{\rm
dec}$ following from the observational constraint $|\fnl|\sim
|f(s)|/r_{\rm dec}\lesssim 10^2$ \cite{Komatsu:2010fb} gets relaxed
in the regions $|f(s)|\ll 1$ which makes it easier to have
$\eta_{\sigma}\gtrsim r_{\rm dec}$. For example, if $|f|\simeq10^{-1}$ then the observational bound on $\fnl$ requires that $r_{\rm dec}\gtrsim 10^{-3}$, while the observed value of the spectral index suggests that $\eta_{\sigma}\lesssim 10^{-2}$, and it could be larger if there is an accidental cancellation between $\eta_{\sigma}$ and $\epsilon$. We therefore conclude that there is a reasonably large parameter space in which Planck may be able to detect both the bispectrum and its scale dependence.


A similar enhancement of scale-dependence could also take place in
other single-source models where isocurvature perturbations of an
initially subdominant field $\chi$ are converted into curvature
perturbations at some later stage, and perturbations of the other fields
can be neglected. The curvature perturbation can be schematically
written in a form analogous to the curvaton case \cite{Alabidi:2010ba}, $\zeta =
r(\zeta_{\chi}+f(s)\zeta_{\chi}^2+\ldots)$, with $r \propto
\rho_{\chi}/\rho$ and $s$ measuring interactions of the $\chi$
field. This yields $\fnl \propto f(s)/r$ where $r$, being independent of the time of horizon crossing of a given mode, does not
contribute to the the scale-dependence. Therefore, $\nfnl = {{\rm
d}\,{\rm ln}|f(s)|}/{{\rm d}\,{\rm ln}\,k} $, and the
scale-dependence gets enhanced in the regions $|f|\ll 1$, provided
they exist. Similar comments apply to the scale-dependence of
$\gnl$.

\subsection{Running of $\nfnl$ and $\ngnl$}

In the regimes of enhanced scale-dependence, $|\nfnl|,|\ngnl|\gg
\eta_{\sigma}$, the first--order derivatives of $\fnl$ and $\gnl$ are
not necessarily enough to describe the scale-dependence but higher--order derivatives may also become relevant. In addition to $\nfnl$
and $\ngnl$, one then needs to consider the running of these
parameters.

Starting from the expressions (\ref{nfnl_single}) and
(\ref{ngnl_single}) it is straightforward to compute the running of
$\nfnl$ and $\ngnl$ \cite{Byrnes:2010xd,Huang:2011di}. The results
can be expressed in the form
  \baq
  \label{afnl}
  \alpha_{\fnl}\equiv\frac{{\rm d}\,\nfnl}{{\rm d}\,{\rm
  ln}\,k}&=&-\nfnl^2+\left(2\epsilon_{\rm H}-\frac{2(n-2)(1+n s)}{2+n (n-1) s}\,\eta_{\sigma}
  \right)\nfnl\ ,\\
  \label{agnl}
  \alpha_{\gnl}\equiv\frac{{\rm d}\,\ngnl}{{\rm d}\,{\rm
  ln}\,k}&=&-\ngnl^2+\left(2\epsilon_{\rm H}-\frac{2(n-2)(n-3-3ns)}{(n-3)(2+n(n-1)s)}\,\eta_{\sigma}\right)\ngnl
  \nonumber \\
  &&-\frac{4n^2(n-2)s}{(n-3)(2+n(n-1)s)}\frac{\fnl^2}{\gnl}\nfnl\eta_{\sigma}\
  ,
  \eaq
where $\epsilon_{\rm H}=-\dot{H}/H^2$.

For $|\nfnl|\gg {\cal O}(\eta_{\sigma})$, equation (\ref{afnl})
gives $\alpha_{\fnl}=-\nfnl^2+{\cal O}(\epsilon)\nfnl$, which
implies that ${\rm d}^2 \fnl/{\rm d}\,{\rm ln}\,k^2 = {\cal
O}(\epsilon){\rm d} \fnl/{\rm d}\,{\rm ln}\,k$. ${\cal O}(\epsilon)$
denotes slow roll corrections proportional to $\epsilon_{\rm H}$ or
$\eta_{\sigma}$. Similarly all higher order derivatives are slow
roll suppressed and $\fnl$ can be expanded around some reference
scale $k_0$ as
  \beq
  \label{fnl_closed}
  \fnl(k)=\fnl(k_0)\left(1+\nfnl(k_0)\,{\ln}\frac{k}{k_0}\left(1+\sum_{n=1}^{\infty}{\cal
  O}(\epsilon^n)\,{\ln}^n\frac{k}{k_0}\right)\right)\ .
  \eeq
If we consider a range of $k$-modes corresponding to a few
$e$-foldings at most, the corrections ${\cal O}(\epsilon){\rm
ln}\,k/k_0$ are tiny and can be neglected. This leaves us with the
compact result
  \beq
  \label{fnl_nonpert}
  \fnl(k)=\fnl(k_0)\left(1+\nfnl(k_0)\,{\ln}\frac{k}{k_0}\right)\ .
  \eeq
Note that this is a non-perturbative expression valid to all orders
in $\nfnl$ and not just a truncated expansion. A similar result can
be derived for $\gnl$ whenever $|\ngnl|\gg \eta_{\sigma}$.

As a curiosity, we notice that expressions formally similar to
(\ref{fnl_nonpert}) appear in models where non-Gaussianity is
generated by classical superhorizon loops \cite{Kumar:2009ge} (see also   \cite{Suyama:2008nt,Bramante:2011zr}).
In such scenarios the logarithm ${\rm ln}\,(k/k_0)$ gets
replaced by ${\rm ln}\,(k L)$ where $L$ is an arbitrary infrared
cut-off scale \cite{Seery:2010kh}. While the role of $L$ is somewhat subtle, our expression is manifestly independent of $k_0$.

\section{Quartic interactions}
\label{sec4}

\begin{figure}[h!]
    \begin{center}
    \includegraphics[width=15 cm, height= 6 cm]{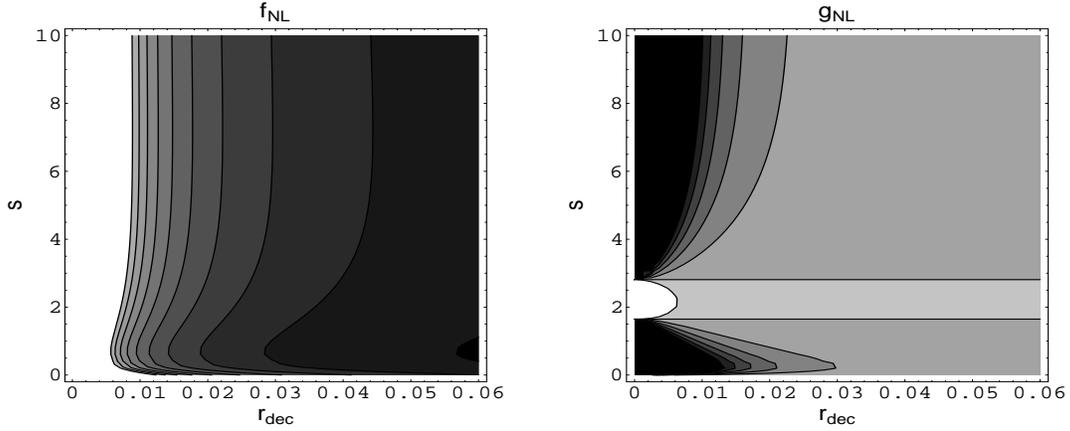}
    \caption{$\fnl$ and $\gnl$ plotted against $r_{\rm dec}$, measuring the curvaton energy density at the time of decay,
    and the self-interaction strength parameter $s$.
    The contours in the left panel run from $10$ (black) to $100$ (white) with a spacing of $10$.
    In the right panel the contours run from $-5000$ (black) to $1000$ (white) with a spacing of
    $1000$; the $0$-contours correspond to the two horizontal lines.}
    \label{fig:fnlgnl}
    \end{center}
  \end{figure}

\begin{figure}[h!]
    \begin{center}
    \includegraphics[width=15 cm, height= 4 cm]{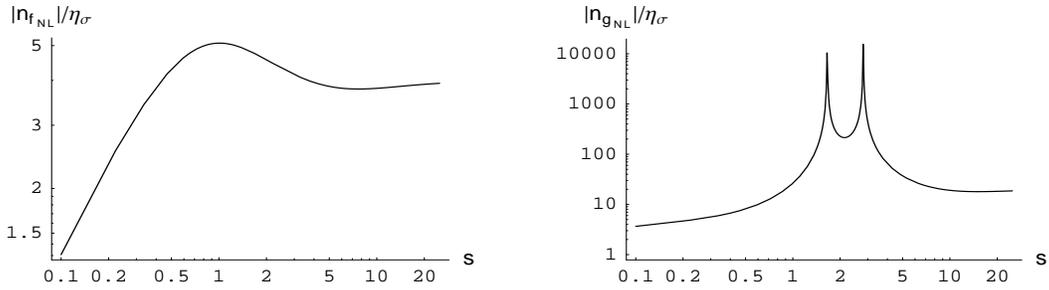}
    \caption{$|\nfnl|/\eta_{\sigma}$ and $|\ngnl|/\eta_{\sigma}$
    plotted as a function of the self-interaction strength parameter $s$ on logarithmic scales.}
    \label{fig:ab}
    \end{center}
  \end{figure}

In this Section we discuss curvaton models with (marginally)
renormalizable four-point interactions
  \beq
  V=\frac{1}{2}m^2\sigma^2+\lambda\sigma^4\ .
  \eeq
In \cite{Enqvist:2009zf} it was shown that for this class of models
$\sigma_{\rm osc}$ can be approximated by,
  \beq
  \label{curvaton_sigmaosc}
  \so\simeq \sigma_{*}\,\frac{1.3\,e^{-0.80 \sqrt{\lambda}\sigma_{*}/m}
  }{|\Gamma(0.75+i\,0.51 \sqrt{\lambda}\sigma_{*}/m)|}=\sigma(t_k)\,\frac{1.3\,e^{-0.56 \sqrt{s}}
  }{|\Gamma(0.75+i\,0.36 \sqrt{s})|}(1+{\cal O}(\eta_{\sigma}))\ ,
  \eeq
where $\sigma_{*}=\sigma(t_{k})(1+{\cal O}(\eta_{\sigma}))$ denotes
the curvaton value at the end of inflation. Using this result, it is
now straightforward to compute the amplitudes $\fnl$ and $\gnl$ and
their scale-dependence $\nfnl$ and $\ngnl$, given by equations
(\ref{fexps})--(\ref{ngnl_single}). The results are depicted in
Figs.~\ref{fig:fnlgnl} and~\ref{fig:ab}.

For the quartic case, $\fnl$ does not show significant oscillatory
features and consequently $\nfnl$ does not get enhanced for any
value of the self-interaction strength parameter $s$. On the other
hand, $\gnl$ changes sign twice around $s\sim 2$, and in this region
$\ngnl$ is considerably enhanced. The two divergent spikes seen in
the plot for $\ngnl$ correspond to the points where $\gnl=0$. When moving
away from these points, $\gnl$ starts to grow while $\ngnl$ still
remains large. In the self-interaction dominated regime $s\gg 1$,
both $\nfnl$ and $\ngnl$ asymptote to constant values. Indeed, in
this limit equation (\ref{curvaton_sigmaosc}) reduces to a simple
power law $\so \propto \sigma(t_k)^{3/4}$ which yields
$\fnl=10/(12r_{\rm dec})\,,~\nfnl = 4\eta_{\sigma} $ and
$\gnl=-25/(54 r_{\rm dec}^2)\,,~\ngnl=-20\eta_{\sigma}$.

The quartic model nicely demonstrates how the interacting curvaton
scenario can generate strongly scale-dependent non-Gaussianity.
However, the results for $\ngnl$ are of limited observational
interest because the bound $|\fnl|\lesssim 10^{2}$ requires $\gnl$
to be small $|\gnl|\lesssim 10^3$, see Fig.~\ref{fig:fnlgnl}. This
is too small to be detectable with the CMB \cite{Smidt:2010ra}.

\section{Non-renormalizable interactions}
\label{sec5}

For non-renormalizable curvaton potentials, $n=6$ and $n=8$ in
(\ref{V}), we have used numerical methods similar to
\cite{Enqvist:2009zf} to study the dynamics and compute the
scale-dependence. In the interaction dominated regime, $s\gg 1$, it
is also possible to obtain simple analytical estimates as we briefly
discuss at the end of this section, see Sec.~\ref{sec:interactiondominated}.

\begin{figure}[htbp]
  \begin{center}
    \resizebox{100mm}{!}{
    \includegraphics{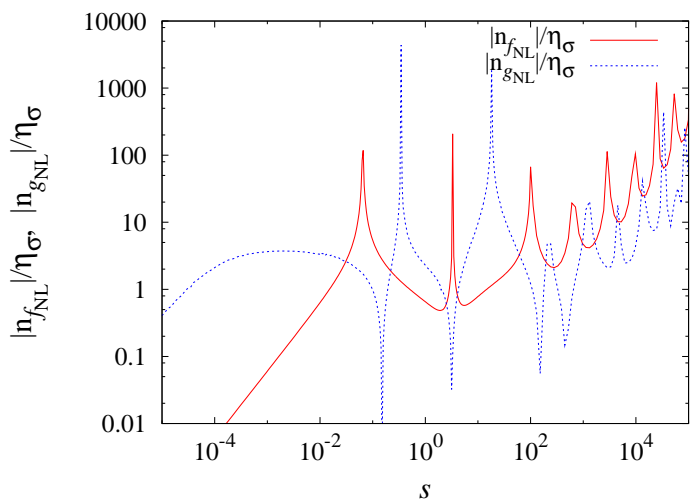}
}
  \end{center}
  \caption{Plot of $|n_{f_{\rm NL}}|/\eta_\sigma$ (red--solid line) and $|n_{g_{\rm NL}}|/\eta_\sigma$ (blue--dashed line) as a function of the self-interaction
strength parameter  $s$ for $n=6$.}\label{fig:nfnl_ngnl_6}
\end{figure}

The presence of non-renormalizable self-interactions renders both
$\fnl$ and $\gnl$ oscillatory functions of the self-interaction
strength parameter $s$ \cite{Enqvist:2009ww}. The oscillatory
behaviour can lead to great enhancement of the scale-dependence, as
discussed above. This is clearly seen in Figure
\ref{fig:nfnl_ngnl_6} which shows $|\nfnl|/\eta_{\sigma}$ and
$|\ngnl|/\eta_{\sigma}$ as a function of $s$ for $n=6$. The results
for $n=8$ are qualitatively similar.

The spikes in the behaviour of $|\nfnl|$ in Figure
\ref{fig:nfnl_ngnl_6} correspond to points where $\fnl$ crosses
zero. In the vicinity of these points $\fnl$ takes non-zero, and for
small $r_{\rm dec}$ observable, values while $|\nfnl|$ is one or two orders
of magnitude enhanced compared to the slow--roll scale
$\eta_{\sigma}$. Similar comments apply to $|\ngnl|$ whose behaviour
is illustrated in the same figure.

\begin{figure}[!h]
\centering \subfigure[$n=4$] {
\includegraphics[width=7 cm]{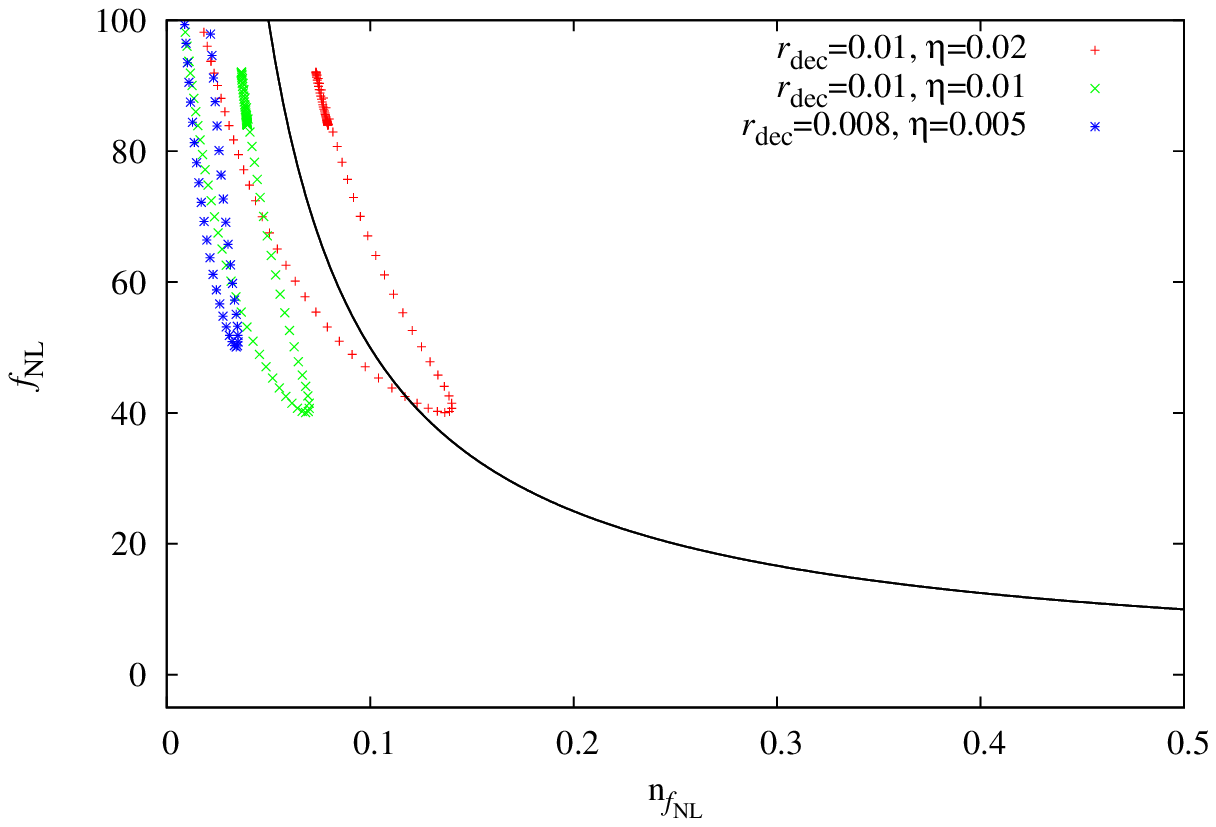}
\label{fig:fnl_nfnl_4}}
\subfigure[$n=6$]{
\includegraphics[width=7 cm]{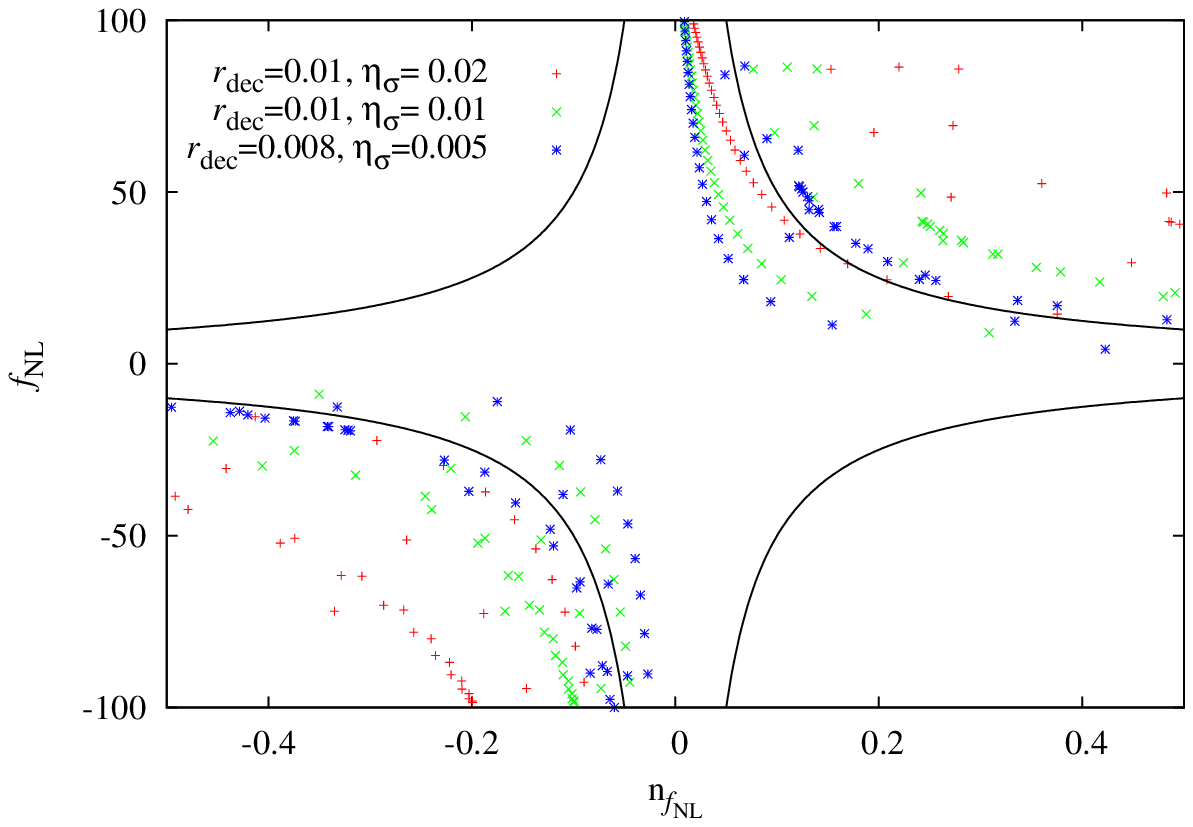}
\label{fig:fnl_nfnl_6}}
\subfigure[$n=8$]{
\includegraphics[width=7 cm]{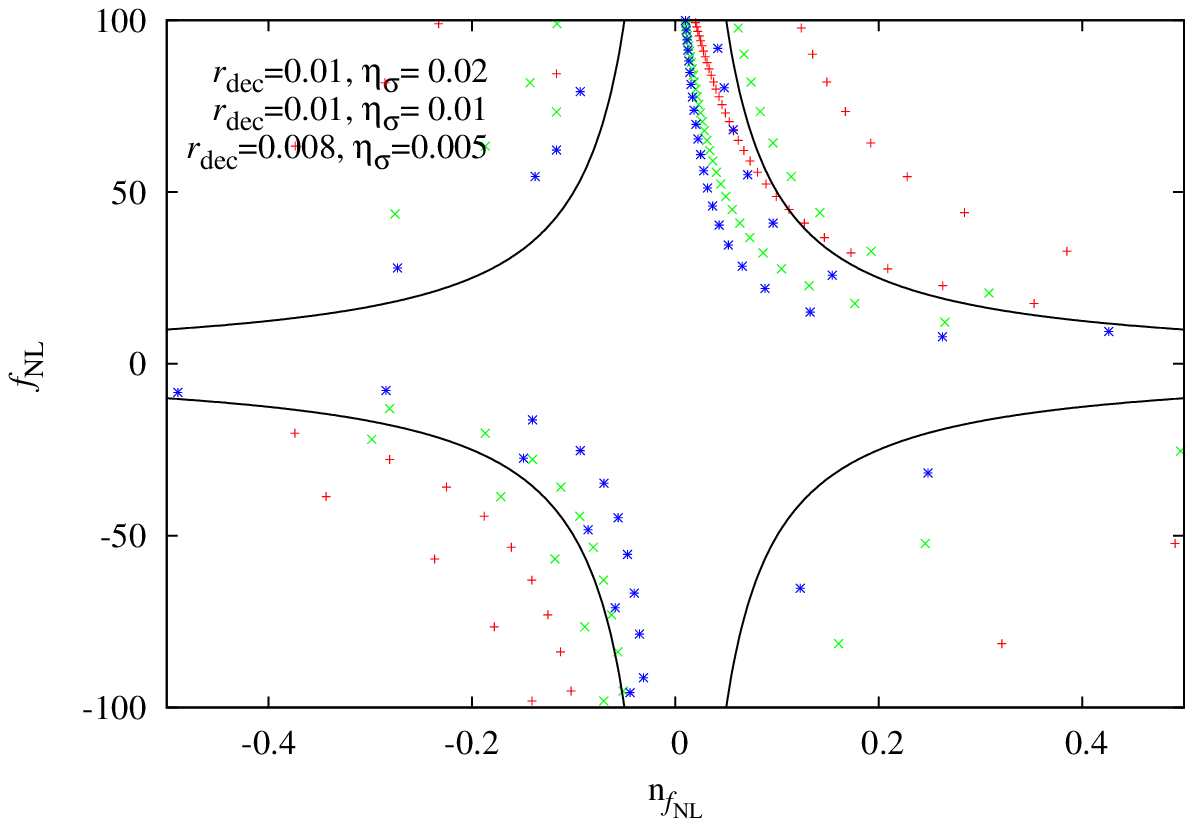}
\label{fig:fnl_nfnl_8}}
\subfigure[$n=6$]{
\includegraphics[width=7 cm]{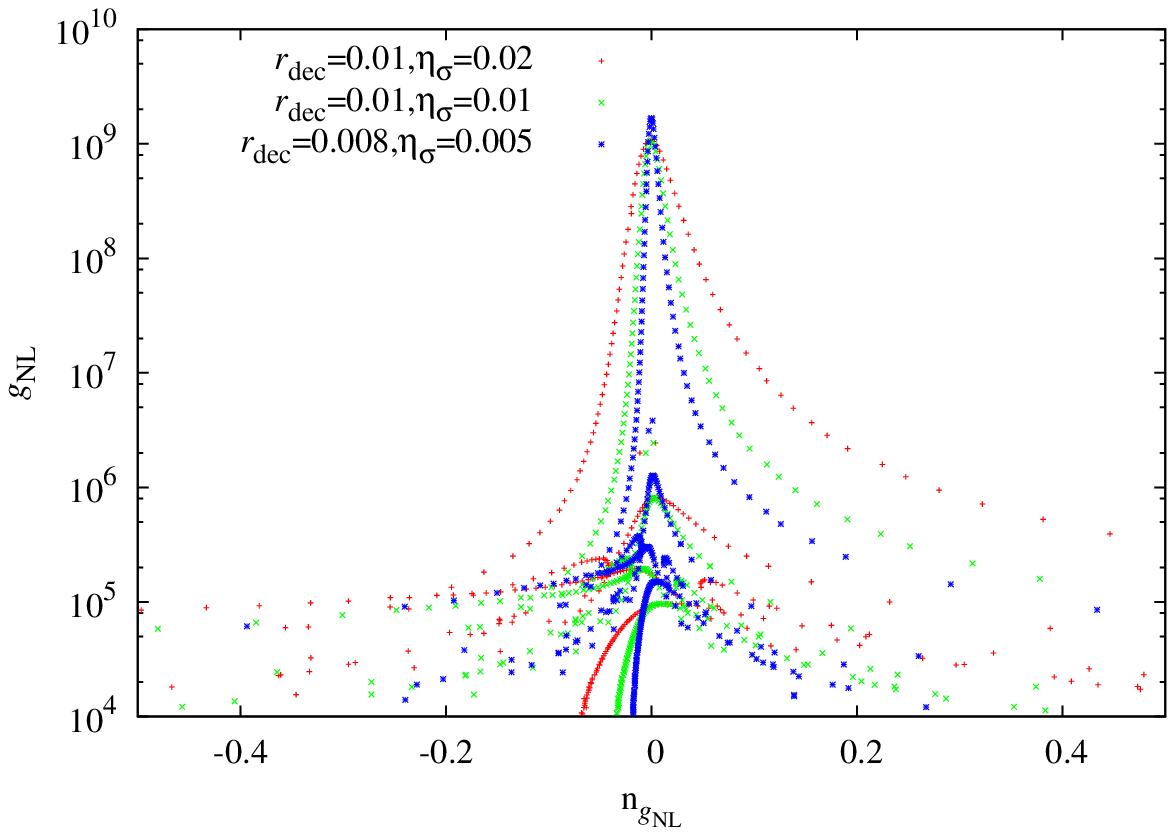}
\label{fig:gnl_ngnl_6}}
\caption{$\fnl$ vs. $\nfnl$ for $n=4,6,8$ and $\gnl$ vs. $\ngnl$ for
$n=6$ with different choices of $r_{\rm dec}$ and $\eta_{\sigma}$.
The points shown in the figures range from $s=10^{-5}$ to
$s=10^{5}$. The black curves depict the forecasted observational sensitivity of
Planck, the region of detectable $\nfnl$ lies outside the curves,
i.e.~further from the origin.}
\label{fig:fnl_nfnl_etc}
\end{figure}

In Figure \ref{fig:fnl_nfnl_etc} we compare our results with the
predicted accuracy of Planck for observing the scale-dependence.
Figures \ref{fig:fnl_nfnl_4}, \ref{fig:fnl_nfnl_6} and
\ref{fig:fnl_nfnl_8} show $\fnl$ against $\nfnl$, scanning from
$s=10^{-5}$ to $s=10^{5}$ and keeping $r_{\rm dec}$ and
$\eta_{\sigma}$ fixed. Outside the black lines which denote $|\nfnl\fnl| =
5$, the scale-dependence is detectable by Planck at the $1$-$\sigma$
level (and at the $2$-$\sigma$ level with CMBPol)
\cite{Sefusatti:2009xu}. For $n=6$ and $n=8$, the points
corresponding to a given choice of $r_{\rm dec}$ and $\eta_{\sigma}$
do not lie on a single curve as in the case $n=4$. This is again a
manifestation of the oscillatory behaviour of $\fnl$, characteristic
for non-renormalizable self-interactions. However, it is noteworthy
that it is possible to generate observable scale-dependence even for
$n=4$, provided that the ratio $\eta_{\sigma}/r_{\rm dec}$ is large
enough.

For comparison, we have also plotted $\gnl$ against $\ngnl$ for
$n=6$ in Figure \ref{fig:gnl_ngnl_6}. As $\gnl$ and $\ngnl$ feel
derivatives up to third order, this plot shows considerably more
structure than the corresponding result for $\fnl$ and $\nfnl$
(Figure \ref{fig:fnl_nfnl_6}), which only feel derivatives up to
second order. There are currently no forecasts on how well $\ngnl$ could be measured.

\begin{figure}[!h]
\centering \subfigure[$n=4$] {
\includegraphics[width=7 cm]{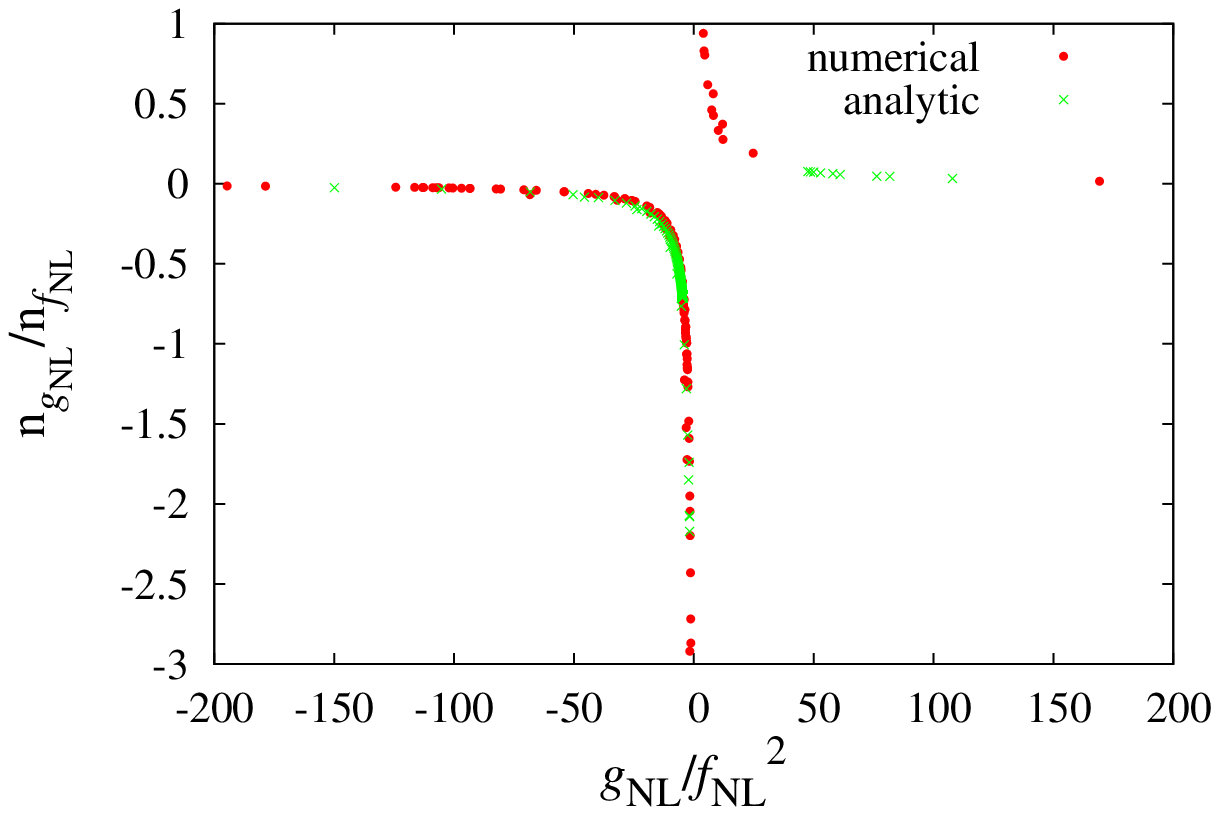}
\label{gf-nfng-nis4}}
\subfigure[$n=6$]{
\includegraphics[width=7 cm]{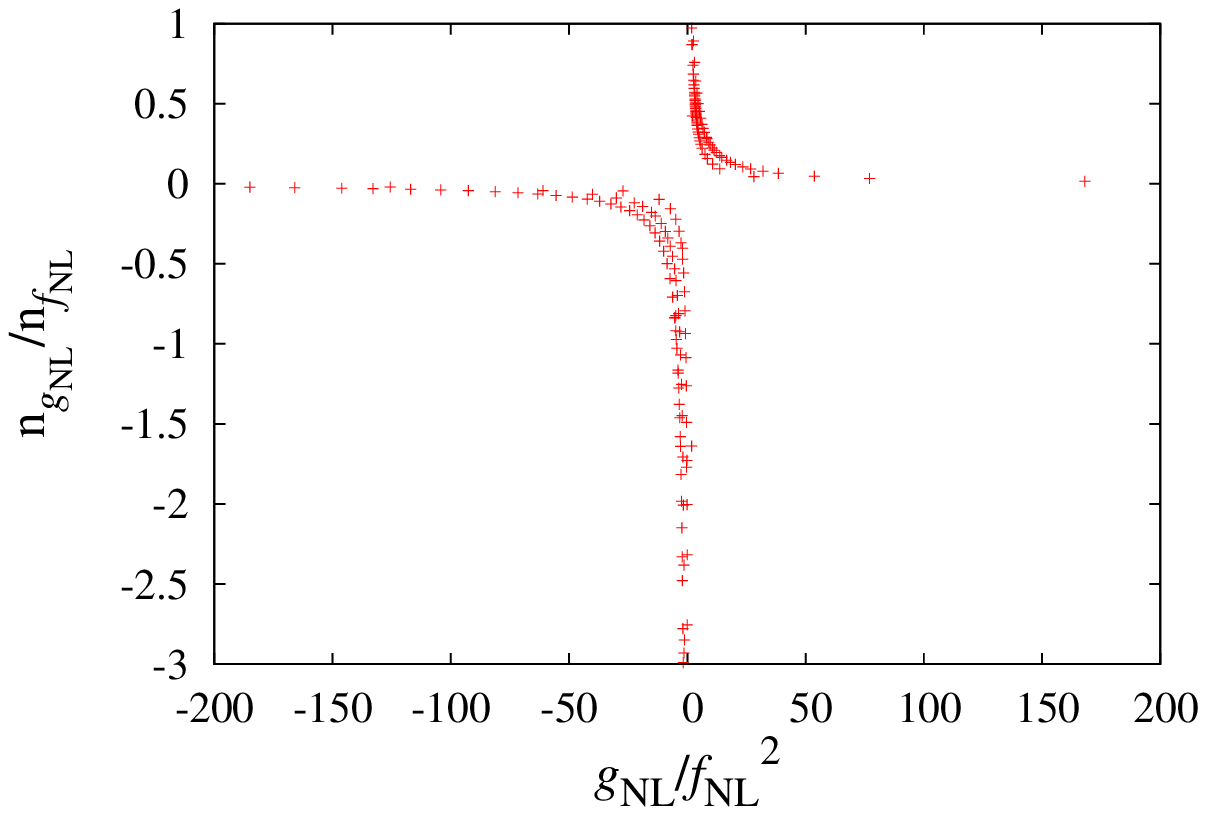}
\label{}}
\subfigure[$n=8$]{
\includegraphics[width=7 cm]{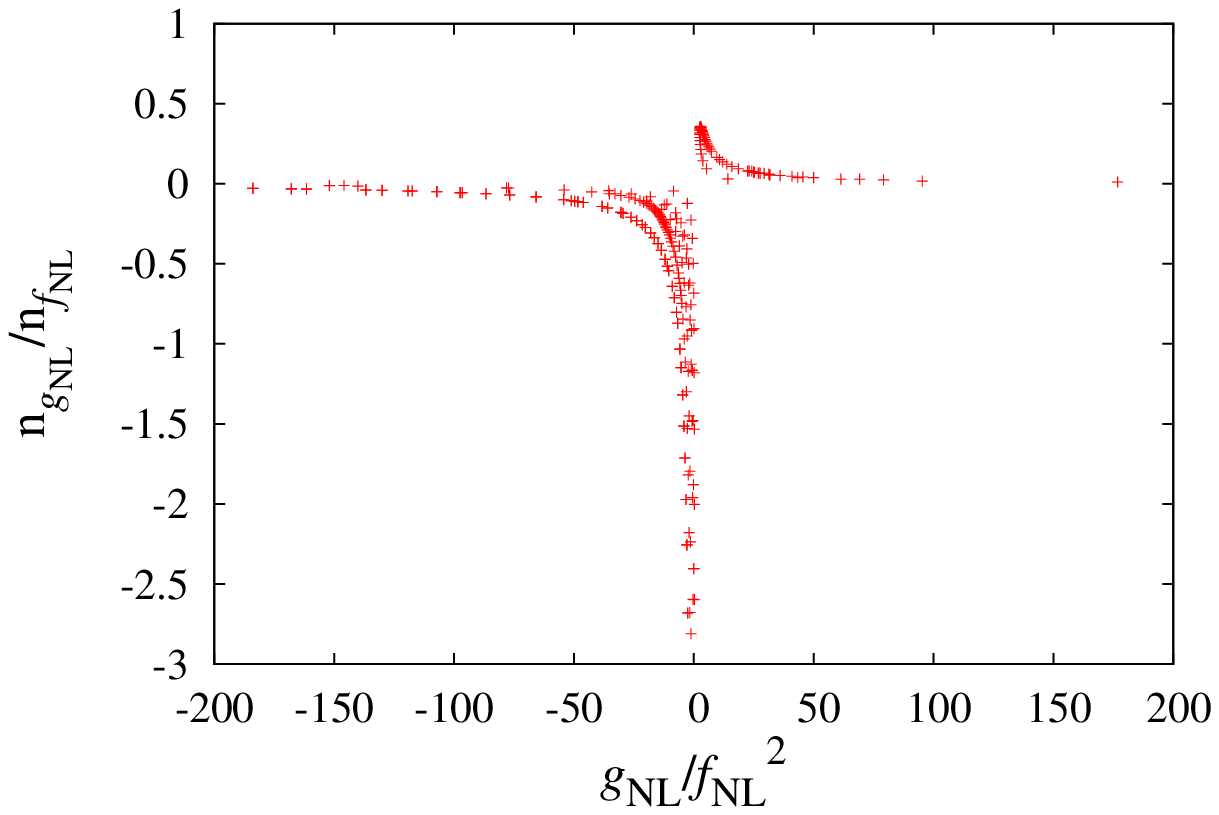}
\label{}}
\caption{The observables $\gnl/\fnl^2$ and $\ngnl/\nfnl$, which
depend on the self-interaction strength $s$ only, plotted for
$s=10^{-5}...10^{5}$. The points outside the curves are not
accessible for any parameter values in the self-interacting curvaton
scenario. In the case of a renormalisable self-interaction, $n=4$,
we also plot points in green calculated using the analytic formula.}
\label{fig:observables}
\end{figure}

In Figure \ref{fig:observables} we plot $\ngnl/\nfnl$ against
$\gnl/\fnl^2$.
These ratios only depend on the self-interaction strength parameter
$s$. The points depicted in the figure range from $s=10^{-5}$ to
$s=10^{5}$. The predictions asymptote to constant values both for
$s\rightarrow 0$ and $s\rightarrow \infty$ and extending the plot
region to smaller or larger $s$-values essentially leaves the plots
unchanged. Therefore, only the points that lie on the curves seen in
Figure \ref{fig:observables} are accessible in the self-interacting
curvaton scenario. The region outside the curves can not be accessed
for any parameter values. Despite its very rich structure and broad
range of different observational imprints
\cite{Enqvist:2009ww,Enqvist:2009zf}, the self-interacting curvaton
scenario could therefore be ruled out by a combined detection of
$\fnl$ and $\gnl$ and their scale-dependencies.

\subsection{Interaction dominated regime}\label{sec:interactiondominated}

In the interaction dominated regime $s\gg 1$, it is possible to derive
analytical results even for the non-renormalizable case. In this
regime the curvaton oscillations start in the non-renormalizable
part of the potential and the transition to the quadratic potential
takes place relatively late after the onset of oscillations. The
dynamics can therefore be described by the simple scaling law
$\rho_{\sigma}\propto a^{-6n/(n+2)}$, unlike for smaller values of
$s$ \cite{Enqvist:2009zf}. For $s\gg 1$, the transition time can be
estimated by $\lambda\sigma^{n-2}\sim m^2$ which yields
${\sigma}_{{\rm osc}} \propto \sigma(t_k) s^{-1/8}(1+{\cal
O}(\eta_{\sigma}))$. Using this in equations (\ref{fexps}) --
(\ref{ngnl_single}) we obtain the results
   \baq
   \label{results_sgg1}
   \fnl&=&\frac{1}{r_{\rm dec}}\frac{5(6-n)}{2(10-n)}+{\cal O}\left(\frac{\eta_{\sigma}}{r_{\rm dec}}\right)
   \ ,\;\; \nfnl\fnl=\frac{\eta_{\sigma}}{r_{\rm dec}}\frac{10(n-2)}{10-n}
   +{\cal O}\left(\frac{\epsilon^2}{r_{\rm dec}}\right)\ ,\\
   \gnl&=&\fnl^2\frac{2(2-n)}{3(6-n)}+{\cal O}\left(\frac{\eta_{\sigma}}{r_{\rm dec}^2}\right)
   \ ,\;\; \ngnl\gnl=\frac{\eta_{\sigma}}{r_{\rm dec}^2}\frac{50(6+n)}{3(10-n)}
   +{\cal O}\left(\frac{\epsilon^2}{r_{\rm dec}}\right)\ .
   \eaq

The results for $n=4$ agree with the discussion in Section
\ref{sec4}. For $n=6$, the amplitudes $\fnl$ and $\gnl$ vanish to
leading order in slow roll, $\fnl={\cal O}(\eta_{\sigma}/r_{\rm
dec})$, $\gnl=(\eta_{\sigma}/r_{\rm dec}^2)$, and we find $\nfnl =
{\cal O}(1), \ngnl = {\cal O}(1)$. The first order slow roll
computation used in our analysis is not enough to derive explicit
results for this case but the level of non-Gaussianity is clearly
unobservably small. For $n=8$ we find $\fnl=-5/(2r_{\rm dec})$,
$\gnl = 2\fnl^2$, $\nfnl=-12\eta_{\sigma}$ and
$\ngnl=28\eta_{\sigma}$. We have checked that these analytical
estimates agree with our numerical simulations.

We note that our results do not agree with those in
\cite{Huang:2008zj}, in particular compare with Eq.~(2.56) of
\cite{Huang:2008zj} in the large $s$ limit, where the sign of $\gnl$
was found to be positive for all values of $n$. We do not attempt to
explain the difference, but we note again that we have found a good
agreement between our analytic and numerical results, both here for the interaction dominated regime as well as for the $n=4$ case, see for
example Fig.~\ref{gf-nfng-nis4}.

\section{Conclusions}
\label{sec6}

We have studied the scale-dependence of the non-linearity parameters, especially of $\fnl$ and $\gnl$ in the curvaton scenario allowing for the possibility of a large self interaction. We have found a rich structure in the results, and that a much larger scale dependence of the non-linearity parameters is possible than may be expected by comparison to the observed spectral index of the power spectrum. This boosts the observational prospects for detecting non-Gaussianity and its scale-dependence and shows that Planck may achieve a simultaneous detection of $\fnl$ and $\nfnl$. This would put stringent constraints on the curvaton scenario and rule out its simplest and most studied version, the curvaton with a quadratic potential.

Although the richness of the results, as shown in the many plots, makes it hard to make firm predictions of the curvaton scenario, one can observe some interesting general trends in the results. In general, increasing the strength of the self interaction $s$, and/or the power of the self-coupling $n$ leads to larger values of the scale-dependence as well as an oscillatory structure. Due to their dependence on a third derivative, $\gnl$ and $\ngnl$ have a more complex structure than $\fnl$ and $\nfnl$, which only depend on a second derivative. In the limit of no self-interaction, i.e.~$s=0$, we recover a constant and potentially large $\fnl$. However $\gnl\simeq0$ is far too small to be observable in this limit.

Previous studies of scale-dependence in non-quadratic curvaton scenario's had focussed on a region with only small self-interactions. In that regime and for the models studied in this article it was found that $\nfnl>0$ \cite{Byrnes:2010xd}, while for an axionic curvaton potential the opposite sign was found \cite{Huang:2010cy}. In both cases the sign of $\nfnl$ was given by the sign of the third derivative of the potential. However we have here shown that even for a fixed potential the sign of $\nfnl$ may oscillate,  depending on the initial field value, which affects the self--interaction strength $s$. For a quartic self--interaction there are only oscillations in the sign of $\ngnl$ but not $\nfnl$, for higher powers of the self interaction multiple oscillations in both of these parameters occurs.

The scale dependence of the non--linearity parameters does linearily depend on the $\eta_{\sigma}$ slow-roll parameter, just as the power spectrum's spectral index does (provided that the $\epsilon$ slow-roll parameter is subdominant, one has $n_s-1=2\eta_{\sigma}$). However the numerical coefficient in the case of $\nfnl$ and $\ngnl$ also depends on the value of the self--interaction strength and can become very large in cases where the amplitude of the non-linearity parameters become supressed, but they may still remain observable provided that the curvaton is sufficiently subdominant at the time of decay. In the limit of a large self--interaction strength with an octic self interaction, which does not correspond to any suppression of the non-linearity parameters we find $\nfnl\simeq -12\eta_{\sigma}$ and $\ngnl\simeq28\eta_{\sigma}$, both of which are an order of magnitude larger than the spectral index.

This rich structure of the self--interacting curvaton does not make the model unpredictive or unfalsifiable, there are model constraints, for example on how late the curvaton decays and on the observed amplitude of the power spectrum which restrict the allowed model parameters. The fact that the model can be observationally ruled out is clear from the plots relating $\fnl, \gnl, \nfnl$ and $\ngnl$ given in Fig.~\ref{fig:observables}, only a few lines in parameter space are allowed.

\acknowledgments{

The authors are grateful to Qing-Guo Huang for useful correspondence. We thank Takeshi Kobayashi for pointing out an incorrect normalisation in Fig.~4 of the first version of this paper. CB thanks Nordita and the University of Helsinki for hospitality
during visits while part of this work was carried out.
TT would also like to thank the University of Helsinki for hospitality
during the visit. CB and SN are
grateful to the ICG, University of Portsmouth for hospitality. KE is
supported by the Academy of Finland grants 218322 and 131454.
The work of TT is partially supported by the Grant-in-Aid for Scientific
research from the Ministry of Education, Science, Sports, and
Culture, Japan, No. 23740195 and Saga University Dean's Grant 2011 For
Promising Young Researchers.

}

\appendix

\section{On the accuracy of the results}\label{appendix}

In computing the scale-dependence, we have neglected the
non-Gaussianities of the curvaton perturbations
$\delta\sigma_{\k}(t_k)$ at horizon crossing, following
\cite{Byrnes:2009pe,Byrnes:2010ft}. Since we assume canonical
slow-roll dynamics for the curvaton during inflation, the neglected
parts are in general slow-roll suppressed. (For a discussion of the
scale-dependence of the three and four-point functions of a test
field $\delta\sigma_{\k}(t_k)$, see \cite{Bernardeau:2010ay}.) These
non-Gaussianities can however play a key role if $|\nfnl|$ or
$|\ngnl|$ become of order unity.

In this Appendix we address this issue by considering a
(unrealistic) toy model with a quartic curvaton
  \beq
  V(\sigma)=\lambda\sigma^4\ ,
  \eeq
and a vanishing classical background field $\sigma=0$. The
inflationary stage is described by a de Sitter solution. While this
model cannot lead to a successful curvaton scenario, it clearly
demonstrates how the non-Gaussianities of $\delta\sigma_{\k}(t_k)$
can become important.

The connected four-point function of the massless curvaton
fluctuations, evaluated at some time $t_i$ after the horizon exit of
all the four modes, can be written as
\cite{Zaldarriaga:2003my,Bernardeau:2003nx,Bernardeau:2010ay}
  \baq
  \label{sigma4point}
  \nonumber
  \langle\delta\sigma_{\k_1}(t_i)\delta\sigma_{\k_2}(t_i)\delta\sigma_{\k_3}(t_i)\delta\sigma_{\k_4}(t_i)\rangle&=&
  (2\pi)^3\delta(\sum\k_{m})\frac{8\lambda}{H^2}\left(\gamma+\xi(\{k_m\})+{\rm ln}\frac{\sum
  k_m}{k_i}\right)\times\\&&
  \left(P(k_1)P(k_2)P(k_3)+{\rm perm.}\right) +{\cal O}(\lambda^2)\ .
  \eaq
Here $\gamma\simeq 0.58$ is the Euler-Mascheroni constant and
$\xi(\{k_m\})$ is a dimensionless function of the all the four
wavenumbers $k_m$, see \cite{Bernardeau:2010ay} for details. $P(k)$
is the spectrum of curvaton fluctuations and $k_i$ is the mode
crossing the horizon at $t_i$. To first order in the coupling
$\lambda$, there are no other connected $n$-point functions $(n>2)$.

The four--point function affects the trispectrum of curvature
perturbation. Using the $\delta N$ formalism together with
(\ref{sigma4point}), we find the non-linearity parameter $\gnl$
given by
  \beq
  \gnl= \frac{25}{54}\frac{N'''}{N'^3}\left(1+ n^{0}_{\gnl}\left(\gamma+\xi({k_m})+{\rm ln}\frac{\sum
  k_m}{k_i}\right)\right)\ .
  \eeq
Here $n^{0}_{\gnl}$ denotes the scale-dependence given by (\ref{ngnl_single}),
  \beq
  n^{0}_{\gnl}=\frac{N'}{N'''}\frac{V''''}{3H^2}\ ,
  \eeq
computed assuming the modes $\delta\sigma_{\k}(t_{k})$ are Gaussian.
(The part proportional to $\nfnl$ in (\ref{ngnl_single}) vanishes as $V'''(0)=0$ here.)

Concentrating, for simplicity, on equilateral configurations
$k_m=k$, and setting $t_i=t_k$, we obtain
  \baq
  \label{gnl_appendix}
  \gnl&=&\frac{25}{54}\frac{N'''}{N'^3}\left(1+n^{0}_{\gnl}(\gamma-\frac{51}{16}+{\rm ln}\,4)\right)\ ,\\
  \label{ngnl_appendix}
  \ngnl&=&\frac{n^{0}_{\gnl}}{1+n^{0}_{\gnl}(\gamma-\frac{51}{16}+{\rm
  ln}\,4)}\ .
  \eaq
The term $n^{0}_{\gnl}(\gamma-51/16+{\rm ln}\,4)\simeq
-1.2\,n^{0}_{\gnl}$ in (\ref{gnl_appendix}) and
(\ref{ngnl_appendix}) arises from the connected four-point function
(\ref{sigma4point}) of curvaton fluctuations, that is from the
non-Gaussianity of $\delta\sigma_{\k}(t_k)$. For $|n^{0}_{\gnl}|\ll
1$, these corrections can be neglected and we recover the results
previously used in this work. However, if $|n^{0}_{\gnl}|\gtrsim 1$
the corrections clearly have a significant effect on both the
amplitude $\gnl$ and its scale-dependence.

It is straightforward to see that the results get modified in a
qualitatively similar manner for models with realistic potential and
a non-vanishing background field $\sigma$,
  \baq
  \fnl&=&\frac{5}{6}\frac{N''}{N'^2}\left(1+{\cal O}(n^{0}_{\fnl})\right)\
,\qquad \nfnl=\frac{n^{0}_{\fnl}}{1+{\cal O}(n^{0}_{\fnl})}\ ,\\
  \gnl&=&\frac{25}{54}\frac{N'''}{N'^3}\left(1+{\cal O}(n^{0}_{\gnl})\right)\
,\qquad \ngnl=\frac{n^{0}_{\gnl}}{1+{\cal O}(n^{0}_{\gnl})}\ .
  \eaq
Therefore, we conclude quite generally that non-Gaussianities of the
curvaton perturbations can be safely neglected if $|n^{0}_{\fnl}|\ll
1$ and $|n^{0}_{\gnl}|\ll 1$. However, if $n^{0}_{\fnl}$ or
$n^{0}_{\gnl}$ become large a more careful analysis is needed, not
only to compute the scale-dependence but also to find the correct
results for $\fnl$ and $\gnl$.

\end{document}